\documentclass{article}
\usepackage[utf8]{inputenc}
\usepackage{amsmath}
\usepackage[numbers,compress]{natbib}
\usepackage{graphicx}
\usepackage[textwidth=4cm]{todonotes}
\usepackage{authblk}
\usepackage{hyperref}

\DeclareMathOperator*{\argmax}{arg\,max}
\numberwithin{equation}{section}

\author[1]{Marcel Montrey}
\author[1,2]{Thomas R. Shultz}
\affil[1]{\small Department of Psychology, McGill University, Montreal, Quebec, Canada}
\affil[2]{\small School of Computer Science, McGill University, Montreal, Quebec, Canada}

\title{The evolution of high-fidelity social learning \\
\small{Post-print: Proceedings of the Royal Society B \url{http://dx.doi.org/10.1098/rspb.2020.0090}}}
\date{}

\begin{document}

\maketitle

\begin{abstract}
    A defining feature of human culture is that knowledge and technology continually improve over time.
    Such cumulative cultural evolution~(CCE) probably depends far more heavily on how reliably information is preserved than on how efficiently it is refined.
    Therefore, one possible reason that CCE appears diminished or absent in other species is that it requires accurate but specialized forms of social learning at which humans are uniquely adept.
    Here, we develop a Bayesian model to contrast the evolution of high-fidelity social learning, which supports CCE, against low-fidelity social learning, which does not.
    We find that high-fidelity transmission evolves when (1) social and (2) individual learning are inexpensive, (3) traits are complex, (4) individual learning is abundant, (5) adaptive problems are difficult and (6) behaviour is flexible.
    Low-fidelity transmission differs in many respects.
	It not only evolves when (2) individual learning is costly and (4) infrequent but also proves more robust when (3) traits are simple and (5) adaptive problems are easy.
	If conditions favouring the evolution of high-fidelity transmission are stricter (3 and 5) or harder to meet (2 and 4), this could explain why social learning is common, but CCE is rare.
\end{abstract}

\section{Introduction}

Humanity's unparalleled cultural and technological sophistication has been widely attributed to our ability to not just share information, but continually build upon it as well~\cite{Boyd1985,Tomasello1999}.
This process, called cumulative cultural evolution~(CCE), has resulted in knowledge and technology that no single generation could produce on its own.
However, despite extensive evidence of culture in a wide range of species~\cite{Whiten2003}, non-human animals have demonstrated only a limited capacity for CCE.
Not only has observational evidence proved scarce and contentious~\cite{Schofield2018}, but experiments have shown that CCE can be surprisingly difficult to evoke even in closely related primates~\cite{Marshall-Pescini2008,Whiten2009}.
While some examples have been elicited in various species~\cite{Mesoudi2018}, these often involve extensive human intervention and remain comparatively modest.
This raises a question that has perplexed biologists, psychologists and anthropologists alike:
what makes humans, if not unique in our capacity for CCE, uniquely adept at producing it?

CCE arises when social learning preserves information between generations, allowing individual learning or lucky errors in transmission to refine it~\cite{Caldwell2018}.
This process probably depends far more heavily on how reliably information is preserved than on how efficiently it is refined, because the more knowledge accumulates, the more there is to rediscover or reinvent when transmission fails.
Theoretical models explicitly support this idea~\cite{Lewis2012} and often find that transmission fidelity must pass a threshold for culture to accumulate~\cite{Enquist2008} (though see cultural attractor theory~\cite{Buskell2017} for an alternative view).
Notably, humans transmit information with exceptionally high fidelity by not only communicating through language, but also imitating more accurately~\cite{Whiten2009} and robustly~\cite{Subiaul2016}, leveraging a more sophisticated theory of mind~\cite{Call2008}, showing natural inclinations toward pedagogy~\cite{Csibra2009} and practicing a far wider range of teaching behaviours~\cite{Burdett2018}.
This has led to the view that CCE relies on accurate but specialized forms of social learning at which humans are particularly adept~\cite{Boyd1996,Tennie2009,Tomasello1999}.

Precisely what social learning mechanisms underlie CCE remains unclear, however.
Researchers have long emphasized the role of imitation (process-copying) and teaching, drawing sharp contrasts with less accurate forms of social learning like emulation (product-copying)~\cite{Boyd1996,Tomasello1999,Tennie2009}.
On this front, transmission chain and laboratory microsociety studies have yielded contradictory results.
Some have found that imitation and emulation both support CCE~\cite{Caldwell2009,Reindl2017,Zwirner2015}, while others suggest that emulation is insufficient~\cite{Derex2013,Wasielewski2014}.
To complicate matters further, studies emphasizing ecological validity have found that even imitation fails to preserve early stone tool manufacturing (knapping) techniques.
Teaching through gesture~\cite{Cataldo2018} or even language~\cite{Morgan2015} may thus be critical to human-like CCE.

Given this empirical ambiguity, it may be useful to draw a functional distinction between high-fidelity social learning that supports CCE and low-fidelity social learning that does not, regardless of what the underlying mechanisms turn out to be.
Bayesian models drawn from work on language evolution have shown how this can be achieved~\cite{Beppu2009,Whalen2014}.
These reveal that when social learning is captured as sampling and inference, it is too low-fidelity for knowledge to accumulate~\cite{Beppu2009}.
However, when social learning is captured as the direct transmission of beliefs~\cite{Beppu2009} or information about those beliefs~\cite{Whalen2014}, it can give rise to CCE.
A Bayesian framework thus delineates between these two types of learning in a mechanism-agnostic way.

Thus far, such models have largely been used to study cultural evolution in transmission chains.
However, they also present an opportunity to address a more fundamental question:
why would biological evolution produce high-fidelity social learning in some species and not others?
Early models showed that CCE cannot explain the evolution of accurate transmission, because CCE would take many generations to pay for this upfront investment~\cite{Boyd1996}.
As a result, much of the CCE literature has taken such transmission for granted and focused on other factors instead, such as demography, social connectedness, transmission biases and filtering of maladaptive traits~\cite{Mesoudi2018}.
Here, we develop a Bayesian model that contrasts the evolution of high- and low-fidelity social learning directly.
Doing so reveals that high-fidelity transmission evolves under different conditions than social learning that spreads culture but does not refine it.

\section{Model}

Consider a population facing an adaptive problem that involves estimating a set of parameters, $\Theta = \{ \theta_1, \theta_2, ..., \theta_x \}$, where each $\theta$ takes some value between $0$ and $1$.
Beliefs about each $\theta$ are encoded as a probability distribution, $p(\theta)$, that describes which values an individual deems likely and which it does not.
For example, if $\Theta$ encapsulates knowledge about constructing a spear, then elements $\theta_1$, $\theta_2$ and $\theta_3$ could represent the spear's ideal length, diameter and center of gravity (where each characteristic is normalized to fall between some minimum plausible value, represented by $\theta = 0$, and some maximum plausible value, represented by $\theta = 1$).
Similarly, $\Theta$ could encode knowledge about knapping, where $\theta_1$ through $\theta_x$ represent the ideal striking platform angle, flaking surface concavity, distance from the edge, amount of force to apply, etc.
Alternatively, $\Theta$ could capture how much time and effort to devote to one food patch ($\theta_1$) as opposed to another ($\theta_2$) and thus encode a foraging strategy.

Learning occurs when beliefs, $p(\theta)$, change in response to new data, $d$, resulting in an updated set of beliefs, $p(\theta|d)$.
This is modelled as Bayesian inference,
\begin{equation}
	p(\theta|d) = \frac{P(d|\theta) p(\theta)}{\int_0^1 P(d|\theta) p(\theta) d\theta},
	\label{Bayes}
\end{equation}
where posterior beliefs, $p(\theta|d)$, are a product of prior beliefs, $p(\theta)$, and the likelihood of observing the data if those priors are true, $P(d|\theta)$.
Bayesian inference thus takes a learner's beliefs and updates them with new data, such that surprising data change beliefs to a greater extent.
The denominator is simply a normalizing term, which ensures that probabilities integrate to $1$.

Beliefs about each $\theta$ follow a beta distribution and data, $d$, consist of either $n$ samples drawn from the environment or $m$ samples drawn from the population.
After learning, individuals select the most plausible value of $\theta$ as their estimate.
This is the posterior distribution's mode,
\begin{equation}
	\hat{\theta}_\text{MAP} = \argmax_\theta p(\theta|d),
\end{equation}
which can be calculated directly from a beta distribution's shape parameters: $\hat{\theta} = (\alpha - 1) / (\alpha + \beta - 2)$.
This makes our model analytically tractable, because it allows us to reason in terms of the data individuals observe rather than the resulting distributions.

Taken together, these estimates shape the individual's trait.
This trait's efficiency is defined by
\begin{equation}
	z = x - \sum_{i=1}^x \left| \theta_i - \hat{\theta}_i \right|,
	\label{eq:z}
\end{equation}
where $\hat{\Theta} = \{ \hat{\theta}_1, \hat{\theta}_2, ..., \hat{\theta}_x \}$ are the individual's estimates after learning and $x$ is the trait's complexity (the set's cardinality).
When estimates lie close to their ideal values, absolute error is minimized and trait efficiency approaches $z = x$.
Conversely, when estimates lie far from their ideal values, error is maximized and trait efficiency is low ($z = 0$ in the extreme case where each $\theta$ and $\hat{\theta}$ take opposite values of $0$ and $1$).

This formulation makes several simplifying assumptions.
First, a trait's maximum efficiency grows linearly with trait complexity ($x$).
We will see later that this assumption can be weakened to include other growth rates (e.g. logarithmic), subject to some constraints.
Second, each trait has a single optimal variant (a unimodal adaptive landscape), which is not necessarily true in complex domains like tools~\cite{Mesoudi2008}.
Third, each parameter is independent, with the ideal value of one $\theta$ having no effect on the ideal value of others.
In reality, such contingencies do occur, for example, in knapping~\cite{Bril2010}.

In our model, priors reflect common intuitions about $\theta$, whose influence diminishes with learning.
These may arise through similarities in genes, ontogeny, previous experience, etc.
For example, if individuals share only weak intuitions about the ideal length of a spear, some novices could make long spears while others make short ones.
Alternatively, if individuals share strong biases about the amount of force to apply when knapping, novices could consistently overestimate this parameter.
In fact, such a pattern has been observed in experiments~\cite{Bril2010}.
We use an asterisk to denote prior estimates, $\hat{\theta}^*$, and trait efficiency, $z^*$.

An adaptive problem's difficulty can be defined as the average distance between a parameter's ideal value and the prior estimate, $f = \frac{1}{x} \sum_{i=1}^x \big| \theta_i - \hat{\theta}^*_i \big|$.
When problems are hard, the optimal trait is unintuitive and a lot of learning is needed.
Conversely, when problems are easy, efficient solutions are obvious, and there is little or nothing to learn.
This could be due to luck, shared relevant experience or even because evolution has yielded an innate adaptive behaviour~\cite{Wakano2004}.

\subsection{Individual learning}

Individual learning involves interacting directly with the environment, through observation, exploration or trial-and-error.
We formalize this as sampling a random variable $X$, where $\mathrm{E}[X] = \theta$.
For example, in foraging, a sample could indicate whether a given food patch was productive or unproductive, such that $X \sim \mathrm{Bernoulli}(\theta$).
Alternatively, in knapping, a sample could indicate the distance from the platform edge that produced a viable flake.
Distances closer to the ideal could be more likely to succeed, such that $X \sim \mathcal{N}(\theta, \sigma^2)$.
Let $n$ be the average number of samples per parameter.
The average individual learner's estimate is thus
\begin{equation}
	\bar{\hat{\theta}}_\text{I} = \frac{\theta n + \hat{\theta}^* v}{n + v},
	\label{eq:theta_i}
\end{equation}
which reflects the combined influence of the environment ($\theta$) and the prior ($\hat{\theta}^*$).
Note that the relative weight placed on the prior, $v > 0$, can be understood as the number of `virtual samples' that would be needed to form that distribution.
Because more genuine samples are needed to overcome stronger priors, $v$ serves as a measure of conservatism.

Each sample comes at some cost, $c \ge 0$, which represents time, energy, opportunity cost, risk of injury or predation, etc.
More sampling yields a more efficient trait, but comes at a greater overall cost, $c n x$.
For example, making three spears gives more insight into the ideal length of a spear than making two would, but requires additional time, effort, material and risk.
The average individual learner's fitness is thus
\begin{equation}
	\bar{\omega}_\text{I} = \omega_0 + \bar{z}_\text{I} - c n x,
	\label{eq:wI}
\end{equation}
where $\omega_0$ represents aspects of fitness unrelated to learning.

In Bayesian inference, each sample improves accuracy less than the preceding one.
Because the per-sample cost ($c$) is invariant, this captures the notion of diminishing returns.
The optimal learning rate, which maximizes expected utility and fitness, is
\begin{equation}
	n = \sqrt{\frac{f v}{c}} - v.
	\label{eq:n}
\end{equation}
Intuitively, individuals learn more when doing so is inexpensive (low $c$) and problems are difficult (high $f$).
Conservatism ($v$) has a more complicated effect.
When individuals are highly conservative, it's not worth collecting many samples, because beliefs barely change with new data.
Likewise, when priors are extremely diffuse, few samples are needed to sway the learner.
Sampling peaks when behaviour is flexible and priors are weak, but not so weak that individuals show no skepticism toward surprising data.

Combining equations~(\ref{eq:z}), (\ref{eq:theta_i}) and (\ref{eq:n}) gives the average individual learner's trait efficiency:
\begin{equation}
	\bar{z}_\text{I} = x (1 - \sqrt{c f v}).
	\label{eq:zI}
\end{equation}
Because $v > 0$, individual learning cannot reliably acquire the optimal trait, $\bar{z}_\text{I} = x$, unless learning is free ($c = 0$) or the initial trait is already optimal ($f = 0$).
If learning is costly and difficult, then individual learning only partially improves the trait and CCE is needed to reliably acquire the ideal variant.

\subsection{Low-fidelity social learning}
In low-fidelity social learning, individuals learn about the environment by observing others' behavioural outcomes.
For example, seeing many long spears but few short ones is indirect evidence that longer spears are more effective.
In reality, behavioural outcomes often fail to accurately reflect beliefs, resulting in incomplete information and errors in inference~\cite{Henrich2004}.
To capture this notion, learners do not sample an estimate directly, but rather a random variable $Y$, where $\mathrm{E}[Y] = \hat{\theta}$.
For instance, if a demonstrator tries to build spears of length $\hat{\theta}$, errors in production may result in some shorter and some longer ones, such that $Y \sim \mathcal{N}(\hat{\theta}, \sigma^2)$.
Let $m$ be the average number of samples per parameter.
The average low-fidelity social learner's estimate is thus
\begin{equation}
	\bar{\hat\theta}_\text{L} = \frac{\bar{\hat{\theta}} m + \hat{\theta}^* v}{m + v},
	\label{eq:theta_l}
\end{equation}
which reflects the combined influence of social information ($\bar{\hat{\theta}}$) and the prior ($\hat{\theta}^*$).
We confirm in supplementary material, \S1.1 that such social learning does not support CCE, because it cannot improve average trait efficiency over time when combined with individual learning.

Each sample comes at some cost, $k \ge 0$, which represents the expenditure and risk involved in surveilling others.
Collecting additional samples allows learners to more faithfully reproduce the average trait, but comes at a higher overall cost, $k m x$.
The average low-fidelity social learner's fitness is thus
\begin{equation}
	\bar{\omega}_\text{L} = \omega_0 + \bar{z}_\text{L} - k m x.
	\label{eq:w_L}
\end{equation}

As in individual learning, sampling yields diminishing returns.
The optimal social learning rate is
\begin{equation}
    m = \sqrt{\frac{v}{k x} \sum_{i=1}^x \left| \bar{\hat{\theta}}_i - \hat{\theta}^*_i \right|} - v,
	\label{eq:m}
\end{equation}
though such learning should be avoided entirely, $m = 0$, if others haven't improved on the initial trait, $\bar{z} \le z^*$.
More effort is devoted to learning when doing so is inexpensive (low $k$) and there is more knowledge to acquire (the summed term is large).

Combining equations~(\ref{eq:z}), (\ref{eq:theta_l}) and (\ref{eq:m}) gives the average low-fidelity social learner's trait efficiency,
\begin{equation}
	\bar{z}_\text{L} = \bar{z} - \sqrt{k v x \sum_{i=1}^x \left| \bar{\hat{\theta}}_i - \hat{\theta}^*_i \right|}.
	\label{eq:zL}
\end{equation}
Such learning cannot reliably preserve others' knowledge, $\bar{z}_\text{L} = \bar{z}$, unless learning is free ($k = 0$) or there is nothing to learn (the summed term is 0).
Otherwise, some knowledge is lost in transmission and supplanted by prior beliefs~\cite{Beppu2009}.

\subsection{High-fidelity social learning}

High-fidelity social learning involves faithfully reproducing an existing trait, which we formalize as copying another individual's estimates.
One way this could happen is if a learner adopts identical underlying beliefs~\cite{Beppu2009}.
For example, language or gesture could convey everything a teacher knows about where to aim blows when knapping.
Alternatively, a learner could adopt beliefs that are merely compatible with the observed trait (i.e. different distributions with the same posterior mode).
For instance, accurately imitating a demonstrator's construction process could yield spears of the same average length, but subtly different beliefs about the relative efficiency of shorter or longer ones.
In either case, the average high-fidelity social learner's estimate is identical to that of the population, $\bar{\hat{\theta}}_\text{H} = \bar{\hat{\theta}}$, as is its trait efficiency, $\bar{z}_\text{H} = \bar{z}$.
We confirm in supplementary material, \S1.2 that such social learning supports CCE, because it can improve average trait efficiency over time when combined with individual learning.

Each parameter individuals copy comes at some cost.
Thus far, we have assumed that social and individual learning rely on the same cognitive mechanisms~\cite{Heyes2012} and that the evolution of social learning primarily reflects changes in attention and motivation.
However, high-fidelity transmission may involve more specialized and cognitively demanding forms of social learning~\cite{Boyd1996,Tomasello1999,Tennie2009}.
For example, if it involves accurate imitation, then it may require specialized neural machinery for parsing and reproducing bodily actions that has undergone significant elaboration in the hominin lineage~\cite{Stout2017,Subiaul2016}.
Alternatively, if it involves human-like teaching, then it may require the capacity for gesture or even language~\cite{Kline2014}.
Though some researchers argue that high-fidelity transmission is as much a product of cultural as of biological evolution~\cite{Heyes2017}, some genetic endowment is clearly needed, even if this consists of a mere `start-up kit' that is later refined through culture~\cite{Heyes2014}.

That being said, the addition of brain tissue is notoriously energetically expensive, particularly during development~\cite{Isler2014}.
The cost of high-fidelity social learning may thus consist of two components:
a dynamic component, $g_\text{d}$, that reflects the expenditure and risk involved in employing such learning;
and a static component, $g_\text{s}$, that reflects the cost of developing and maintaining it.
This gives an overall cost $g_\text{d} x + g_\text{s} \ge 0$, where the dynamic cost grows with how extensively learning is employed ($x$), but the static cost is invariant.
To capture both components as a single per-parameter cost, we define $g = g_\text{d} + g_\text{s} / x$.
The average high-fidelity social learner's fitness is thus
\begin{equation}
    \bar{\omega}_\text{H} = \omega_0 + \bar{z} - g x.
\end{equation}

\section{Results}
To contrast the evolution of high- and low-fidelity social learning, we track the fate of rare social learning mutants in a monomorphic population of individual learners, where $\bar{z} = \bar{z}_\text{I}$ and $\bar{\omega} = \bar{\omega}_{\text{I}}$.
Social learning goes extinct if these mutants' average fitness ($\bar{\omega}_{\text{L}}$ or $\bar{\omega}_{\text{H}}$) falls below that of the resident type ($\bar{\omega}_{\text{I}}$).
Conversely, social learning evolves if these mutants have higher fitness and their invasion results in either fixation or coexistence (a dimorphic equilibrium).
We do not consider dimorphic resident populations, because for our purposes the effects would be fairly straightforward.
Namely, the resident population's average trait efficiency ($\bar{z}$) would decrease as low-fidelity social learning became more common, which is equivalent to a monomorphic case where high-fidelity transmission is more costly (i.e. $g$ is $\Delta \bar{z} / x$ higher).

While social learning is often subject to frequency-dependent selection~\cite{Rogers1988}, this does not concern us for two reasons.
First, high-fidelity social learning's fitness is not frequency-dependent at all, because it simply maintains the population's average trait ($\bar{z}_\text{H} = \bar{z}$) and this trait's efficiency does not change over time (cf.~\cite{Rogers1988}).
Any fitness advantage it has as a rare mutant thus persists until fixation.
Second, while low-fidelity social learning's fitness is frequency-dependent, this never brings about its extinction.
Rather, as such mutants become more common, their average trait efficiency declines until their fitness equalizes with that of the resident type, $\bar{\omega}_\text{L} = \bar{\omega}_\text{I}$, resulting in a dimorphic equilibrium.
We show in supplementary material, \S2 that such an equilibrium exists and is stable whenever they invade.

Social learning cost is central to our analysis, because it reveals both when a given type of learning could conceivably pay for itself and when it is best equipped to do so.
Setting $\bar{\omega}_\text{L} = \bar{\omega}_\text{I}$ gives the maximum per-sample cost of low-fidelity social learning,
\begin{equation}
    k_{\max} = \frac{\left(\sqrt{c v}-\sqrt{f}\right) \left(2 \sqrt{f-\sqrt{c f v}}+\sqrt{c v}-2 \sqrt{f}\right)}{v},
\end{equation}
and setting $\bar{\omega}_\text{H} = \bar{\omega}_\text{I}$ gives the maximum per-parameter cost of high-fidelity social learning,
\begin{equation}
    g_{\max} = \sqrt{c f v} - c v.
\end{equation}
At or above these values, such learning no longer confers a fitness advantage.
Identifying when $k_{\max} > 0$ or $g_{\max} > 0$ thus reveals the minimum requirements for social learning to evolve.
More importantly, conditions that maximize $k_{\max}$ or $g_{\max}$ reveal when such learning withstands the broadest possible range of costs and is thus most likely to evolve (though such conditions do not necessarily maximize its prevalence in the population, learning rate, etc.).

\subsection{Social learning cost ($k$ and $g$)}
For social learning to evolve, it must either improve on the average trait or reduce the cost of acquiring it~\cite{McElreath2010}.
Although transmission errors can yield a superior trait, lucky mistakes are no more likely to be observed than unlucky ones (cf.~\cite{Henrich2004}).
Therefore, social learning must reduce cost.
Setting $\bar{\omega}_\text{L} > \bar{\omega}_\text{I}$ reveals that low-fidelity social learning evolves when its savings in cost exceed its average loss in trait efficiency
\begin{equation}
	c n x - k m x > \bar{z}_{\text{I}} - \bar{z}_{\text{L}}.
	\label{eq:l-invade}
\end{equation}
The more errors in transmission, the larger the necessary savings.
By contrast, high-fidelity social learning makes virtually no errors in transmission.
It thus evolves ($\bar{\omega}_\text{H} > \bar{\omega}_\text{I}$) when it offers nearly any savings in cost
\begin{equation}
    c n x - g x > 0.
	\label{eq:h-invade}
\end{equation}
(In reality, there will likely always be some slight, non-zero level of error to overcome.)
Taken together, equations~(\ref{eq:l-invade}) and (\ref{eq:h-invade}) imply that high-fidelity social learning tolerates a greater overall cost by maintaining more efficient traits.

\subsection{Trait complexity ($x$)}
Trait complexity can be eliminated from equation~(\ref{eq:l-invade}), because each term grows linearly with $x$.
Doing so yields the equivalent expression $c n - k m > \big| \bar{\hat{\theta}}_{\text{I}} - \bar{\hat{\theta}}_{\text{L}} \big|$, which implies that low-fidelity social learning is as likely to evolve when traits are simple as when they are complex.
The same is not true of equation~(\ref{eq:h-invade}), once we break cost $g$ down into its static and dynamic components.
Instead, the evolution of high-fidelity social learning requires crossing a threshold in trait complexity, $x > g_\text{s} / (c n - g_\text{d})$, which increases with the cost of both having ($g_\text{s}$) and employing ($g_\text{d}$) such learning.

Note that this result is not contingent on our assumption that trait efficiency and learning cost grow linearly with respect to trait complexity, but rather that they grow at the same rate.
For example, $x$ could still be eliminated from equation~(\ref{eq:l-invade}) if efficiency and cost both grew logarithmically (e.g. if increased complexity yielded diminishing returns in efficiency, but learning one parameter made it easier to learn others).

\subsection{Individual learning rate ($n)$}
\begin{figure}
    \centering
    \includegraphics[width=\textwidth]{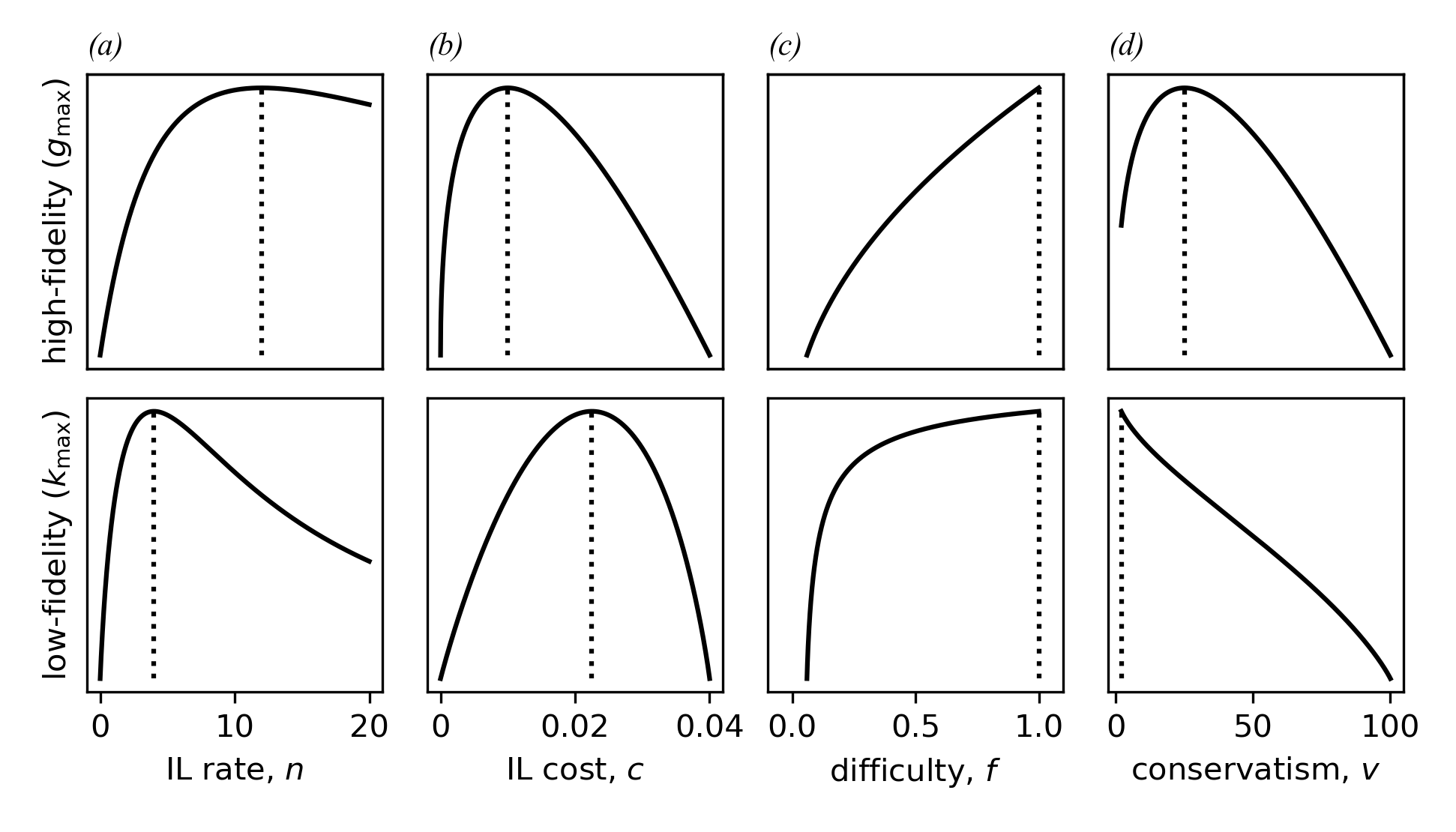}
    \caption{Resilience of high- and low-fidelity social learning as a function of: \textit{(a)} the individual learning rate, \textit{(b)} individual learning cost, \textit{(c)} problem difficulty and \textit{(d)} conservatism. Dotted lines indicate when social learning is most likely to evolve. High-fidelity transmission evolves when individual learning is comparatively \textit{(a)} plentiful and \textit{(b)} inexpensive. Its evolution may also depend on confronting particularly \textit{(c)} challenging adaptive problems, because it accrues the benefits of increased problem difficulty more slowly. Finally, while all social learning benefits from \textit{(d)} behavioural flexibility, high-fidelity transmission could benefit from higher levels of conservatism if these stimulate rather than depress individual learning. Parameters: $c = 0.005$, $f = 0.5$, $v = 12$.}
    \label{fig}
\end{figure}

Social learning can only evolve ($k_{\max} > 0$ or $g_{\max} > 0$) when there is knowledge to acquire, $n > 0$.
However, different types of social learning benefit from vastly different individual learning rates (figure~\ref{fig}\textit{a}).
This can be seen by finding the values of $n$ that maximize $k_{\max}$ and $g_{\max}$ (after first simplifying these expressions by using equation~(\ref{eq:n}) to substitute $c = f v / (n + v)^2$).
Doing so reveals that low-fidelity social learning is most likely to evolve when the individual learning rate is low, $n = v/3$, and beliefs are driven mostly by prior expectations.
By contrast, high-fidelity transmission is most likely to evolve when the individual learning rate is much higher, $n = v$.

\subsection{Individual learning cost ($c$)}
Social learning cannot evolve when individual learning is free, $c = 0$, because it confers no savings.
Similarly, it cannot evolve when individual learning is too expensive to engage in, $c \ge f / v$, because there is nothing to learn.
Between these two extremes, however, different individual learning costs favour different types of social learning (figure~\ref{fig}\textit{b}).
Low-fidelity transmission is most likely to evolve when individual learning is relatively expensive, $c = 9f/(16v)$, whereas high-fidelity transmission benefits from much cheaper individual learning, $c= f/(4v)$.
In fact, the latter regime represents a $5 / 9 \approx 56\%$ reduction in cost.

\subsection{Problem difficulty ($f$)}
Adaptive problems must be sufficiently difficult, $f > c v$, for learning to evolve.
Below this threshold, learning is not cost-effective, because the optimal trait is highly intuitive.
Harder problems favour social learning in particular, which becomes more resilient as difficulty increases: $\partial k_{\max}/\partial f > 0$ and $\partial g_{\max} / \partial f > 0$.
Social learning is thus most likely to evolve when problems are as difficult as possible, $f=1$.
That being said, high- and low-fidelity transmission react differently to increases in difficulty (figure~\ref{fig}\textit{c}).
Normalizing $k_{\max}$ and $g_{\max}$ by their maximum values reveals that $k_{\max} / \left. k_{\max} \right|_{f = 1} > g_{\max} / \left. g_{\max} \right|_{f = 1}$ over $c v < f < 1$.
In other words, low-fidelity transmission accrues the benefits of increased problem difficulty sooner.
Larger increases are thus needed for high-fidelity transmission to reap comparable rewards (i.e. a proportional increase in resilience against cost).

\subsection{Conservatism ($v$)}
Learning can only evolve when the level of conservatism falls below $v < f / c$.
Stronger priors make learning uneconomical, because updating beliefs involves collecting too much data.
Low-fidelity transmission always benefits from reduced conservatism, $\partial k_{\max} / \partial v < 0$, and is thus most likely to evolve when priors are as diffuse as possible (low $v$).
Although high-fidelity transmission also benefits from behavioural flexibility (figure~\ref{fig}\textit{d}), its ideal level of conservatism is somewhat higher, $v = f / (4c)$.
This value is ideal because it maximizes the individual learning rate.

\section{Discussion}
A longstanding question about CCE is why humans acquired this capacity, which appears diminished or absent in other species.
Given the importance of transmission fidelity~\cite{Lewis2012}, one explanation is that CCE relies on powerful but specialized forms of social learning at which humans are uniquely adept~\cite{Boyd1996,Tomasello1999,Tennie2009}.
By characterizing social learning in terms of its ability to support CCE rather than specific underlying mechanisms, we find that high-fidelity transmission evolves under different conditions than less accurate social learning.
Specifically, high-fidelity transmission is most likely to evolve when:
(1) social and
(2) individual learning are inexpensive,
(3) traits are complex,
(4) individual learning rates and
(5) problem difficulty are high, and
(6) behaviour is flexible.
Low-fidelity transmission differs in many respects.
Not only is it most likely to evolve when individual learning is (2) costly and (4) infrequent, but it is also more robust when (3) traits are simple and (5) problems are easy.
If conditions favouring the evolution of high-fidelity transmission are stricter (3 and 5) or harder to meet (2 and 4), this could explain why social learning is common across species, but CCE is rare.

Comparative analyses suggest that reliance on social learning covaries with brain size in primates~\cite{Reader2002,Reader2011}.
Because the hominin brain has undergone several large evolutionary expansions~\cite{Stout2011}, high-fidelity social learning may require the addition of costly brain tissue~\cite{Boyd1996}.
Our model suggests that one way to compensate for this increased expenditure would be to lower other costs associated with social learning.
This could be achieved in several ways.
First, social tolerance and grouping could provide easier, safer and more frequent opportunities to learn from others.
In support of this view, sociability has been found to covary with reliance on social learning both within humans~\cite{Morgan2012} and across primates~\cite{Reader2002}.
Second, extended juvenile periods could free up time for social learning~\cite{Burkart2008} without forgoing the opportunities in reproduction and resource acquisition available to an adult.
Third, proactive prosociality could promote teaching~\cite{Burkart2016}.
Teaching, in this case, does not necessarily refer to the varied and cognitively complex forms it takes in humans~\cite{Burdett2018,Kline2014}, but rather to any instance where individuals modify their behaviour to foster others' learning~\cite{Burkart2016}.
Pedagogy could thus drive its own evolution, with more elaborate forms of teaching evolving in response to this reduction in cost.

Another way to offset the added cost of high-fidelity transmission would be through higher intake~\cite{Isler2014}.
In line with previous models, we find that accurate social learning tolerates a greater overall cost precisely because it yields more efficient traits~\cite{Boyd1996}.
We build on this insight by allowing trait efficiency to grow with trait complexity.
Though this relationship is not universal (e.g. simplifying a trait could make it more efficient), complexity is often indicative of improvement.
For example, as knapping techniques became more elaborate and hierarchically structured, this resulted in better tools~\cite{Stout2010}.
Following this assumption, we find that high-fidelity social learning is more likely to evolve when traits are complex, because the payoffs in trait efficiency dwarf the cost of developing and maintaining such learning.
Unlike other species, early hominins may have crossed a threshold in trait complexity that allowed accurate transmission to evolve.
This initial complexity may have arisen for reasons other than social learning, for example because encephalization allowed for more sophisticated action sequences~\cite{Whiten2003}.

This explanation is consistent with the archaeological record.
Stout and Hecht~\cite{Stout2017} note that the first stone tools (3.3 Ma) saw only intermittent use and that even the early Oldowan technocomplex (2.6--2.0 Ma) gives the impression of being at the limits of hominin ability.
Though the existence of local traditions suggests that Oldowan techniques were culturally transmitted~\cite{Stout2010}, there is a conspicuous lack of evidence for CCE until much later on~\cite{Tennie2016}, following significant increases in brain size~\cite{Stout2011}.
During this early period (and perhaps considerably beyond it~\cite{VanSchaik2019}), social learning seems to have spread and maintained but not significantly refined the manufacture of tools.
Not only is there no clear evidence of high-fidelity transmission~\cite{Tennie2016} but the observed cultural dynamics closely align with those found in our model when individual and low-fidelity social learning are combined (supplementary material, \S1.1).
Namely, a steady state emerges where average trait efficiency remains stable, but knowledge is repeatedly lost and rediscovered (socially mediated serial reinnovation~\cite{Bandini2017}).
In short, rather than high-fidelity social learning spreading and maintaining early lithic technologies, their relative complexity may have instead facilitated its evolution.

The putatively high cost of accurate transmission is only one of the potential impediments to its evolution.
Theory suggests that low individual learning rates could also play a role~\cite{McElreath2010}.
In line with this view, we find that much higher rates may be needed for the evolution of high- rather than low-fidelity social learning.
Notably, the hominin lineage is characterized by large brains and high general intelligence, both of which are predictive of innovation rates in primates~\cite{Reader2002}.
If few species are sufficiently prolific individual learners, this could explain why accurate transmission is rare.

Of course, this raises the question of how adequate individual learning rates could be achieved in the first place.
The most obvious way to stimulate individual learning is to reduce its cost.
Previous theory~\cite{Boyd1985} and experiments~\cite{Morgan2012} warn that doing so can undercut social learning, however.
While we find support for this view, we also find that high-fidelity transmission nevertheless benefits from such reductions.
In practice, many of the same factors that mitigate the cost of social learning could do so for individual learning as well.
First, grouping could reduce the cost of exploration by allowing individuals to diffuse the associated risks~\cite{Moretti2015}.
Second, extended juvenile periods could offer more time for not just social learning, but individual learning as well~\cite{Burkart2008}.
Costs borne by juveniles in protected environments, where others provide food, shelter and predator detection~\cite{Burkart2008}, would be especially affected.
Finally, even teaching could play a role in the form of opportunity scaffolding, where a teacher does not necessarily demonstrate a behaviour, but rather furnishes students with easy and safe opportunities to learn on their own~\cite{Boyette2018}.

Another way to promote individual learning is by facing more challenging adaptive problems.
We find that the evolution of high-fidelity social learning may involve confronting particularly difficult social, ecological and technological challenges (i.e. problems where optimal traits fall far outside the `zone of latent solutions'~\cite{Tennie2009}).
There are several reasons to think that hominins confronted such problems.
First, because bipedalism allows hominins to cover far larger geographical ranges than other primates, with lifetime home ranges several orders of magnitude greater than those of chimpanzees~\cite{Hill2009}, individuals were likely subjected to greater variability in environmental conditions, available resources, potential threats, etc.
If behaviour that is adaptive in one setting is non-adaptive in others, then problems may more frequently require unintuitive solutions.
Second, an exceptionally large proportion of the hominin diet consists of high-quality foods~\cite{Kaplan2000}, such as those procured through hunting, extractive foraging and confrontational scavenging~\cite{Isler2014}.
Compared to foods consumed more regularly by other primates, these are skill intensive and difficult to obtain~\cite{Kaplan2000}.
Finally, new ways of thinking, interacting with others and leveraging technology undoubtedly presented novel problems of their own.
This probably resulted in a unique and challenging cognitive, cultural and technological niche, which further shaped the course of our evolution~\cite{Stout2017}.

Lastly, it is worth commenting on the role of conservatism.
A striking empirical finding is that chimpanzees suffer from remarkable functional fixedness and behavioural conservatism, which are thought to contribute to the paucity of CCE in this species~\cite{Marshall-Pescini2008,Whiten2009}.
We find that conservatism impedes CCE insofar as it disfavours investment into social learning.
However, we also find that high-fidelity social learning could benefit from higher levels of conservatism if these stimulate rather than depress individual learning.
For conservatism to impede the evolution of accurate transmission in particular, some additional assumption must be invoked, namely that such transmission also happens to be comparatively expensive.

Individually, our criteria for evolving improved fidelity of transmission seem simple:
mitigating the cost of learning, confronting harder adaptive problems, acquiring more complex traits, etc.
However, our model emphasizes that meeting any one of these criteria is not necessarily sufficient.
For example, even if migration exposes individuals to less intuitive problems, learning could still be too expensive.
Similarly, even if grouping lowers the cost of learning, traits could still be too simple.
In short, humans probably evolved high-fidelity social learning not by meeting any one (or more) of these criteria perfectly, but by meeting all of them well enough.

\subsubsection*{Authors' contributions}
M.M. and T.R.S conceived the project.
M.M. developed the model, performed the analysis and wrote the manuscript.
T.R.S provided manuscript revisions.

\subsubsection*{Competing interests}
We declare we have no competing interests.

\subsubsection*{Funding}
This work was supported by grants to M.M. (CGS-M and CGS-D) from the Natural Sciences and Engineering Research Council of Canada.

\bibliographystyle{ieeetr}
\bibliography{references}

\end{document}